\documentclass{article}
\usepackage[T1]{fontenc} 
\usepackage[utf8]{inputenc} 
\usepackage{ismir,amsmath,cite,url}
\usepackage{graphicx}
\usepackage{color}
\def\lbd{}	

\usepackage{mdframed}

\usepackage{lineno}
\usepackage{listings}
\usepackage{comment}

\usepackage[bookmarks=false]{hyperref}

\title{Retrieval Augmented Generation of Symbolic Music with LLMs}

\oneauthor
  {Nicolas Jonason, Luca Casini, Carl Thomé, Bob L. T. Sturm} {Division of Speech, Music and Hearing, \\
KTH Royal Institute of Technology, Stockholm \\ {\tt njona@kth.se, casini@kth.se, cthome@kth.se, bobs@kth.se}}

\def\authorname{N. Jonason, L. Casini, Carl Thomé and B.L.T. Sturm}

\begin{document}

\maketitle
\begin{abstract}
We explore the use of large language models (LLMs) for music generation using a retrieval system to select relevant examples. We find promising initial results for music generation in a dialogue with the user, especially considering the ease with which such a system can be implemented. The code is available online.\footnote{\href{https://github.com/erl-j/scrap}{https://github.com/erl-j/scrap}}
\end{abstract}

\section{Introduction}
Large language models (LLMs) are big machine learning models trained on  massive text datasets. 
LLMs have shown to be able to address a wide range of problems if given clear instructions and/or a set of well-curated examples of the solved task \cite{brown2020language}.

Retrieval augmented generation (RAG) is a technique where a language model is augmented with a retrieval system capable of accessing an database.\cite{lewis2021retrievalaugmented}  Liu et al. \cite{liu2021makes} propose using a retrieval system to select a relevant set of few-shot examples for prompting the LLM into solving a particular task instance \cite{lewis2020retrieval}.
Retrieval systems has been integrated in music generation systems before the popularisation of LLMs \cite{zhao2021accomontage, lv2023recreation}.

LLMs have been used to generate music before. Notably, Bubeck et al. \cite{bubeck2023sparks} find that \texttt{GPT-4} \cite{openai2023gpt4} seems to possess some musical notions, and can compose and edit abc-notated\footnote{\href{https://abcnotation.com/wiki/abc:standard:v2.1}{https://abcnotation.com/wiki/abc:standard:v2.1}} music in a zero-shot setting, albeit with trivial harmony. 
Zhang et al. fine-tune variants of GPT-3 on drum tab continuation with promising results \cite{zhang2023llmdrummers}.


In this work we present experiments with retrieval augmented generation of symbolic music within the domain of folk tunes in abc-notation. 

\section{System}
\begin{figure}[hb]
    \centering
    \includegraphics[width=0.925\columnwidth]{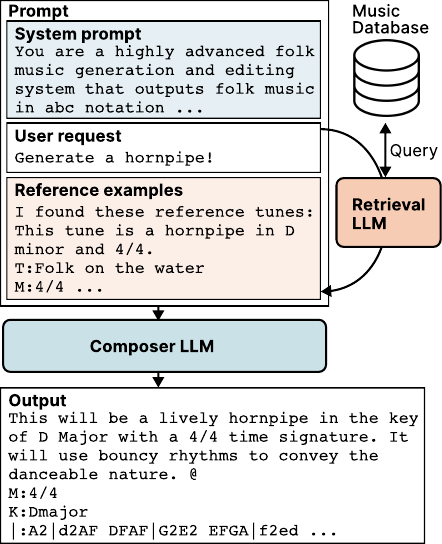}
    \caption{Overview of the system}
    \label{fig:scrap}
\end{figure}

The system is shown in Figure \ref{fig:scrap}.
First, the prompt is initialized with system instructions for the \textit{Composer LLM}. 
For each new user request, the \textit{Retrieval LLM} looks at the user request and formulates a query to a database of symbolic music.
Both the user request and the selected few-shot examples are then appended to the prompt. 
The prompt is then passed to the \textit{Composer LLM} which outputs a new message containing generated symbolic music. 
This process can then be repeated if the user wants to make changes to the generated symbolic music.

\noindent \textbf{Music Database \& Retrieval LLM}
The role of the \textit{Retrieval LLM} is to take the user request and search a database for relevant music examples to add to the prompt. 
Our database consists of a collection of abc-notated Irish traditional folk tunes from thesession.org with tags referring to tune type, mode, and meter. 
While many retrieval approaches have been proposed in the context of RAG, we use a simple tag-based approach as a proof of concept. For each request, the retrieval LLM first returns a set of tags. 
For example, for a request like \texttt{"Give me a lively irish-sounding tune in G dorian"}, the \textit{Retrieval LLM} might generate the tags \texttt{\{jig, reel, dorian\}}. 
The most relevant examples are selected according to Jaccard similarity \cite{jaccard} to the set of tags extracted from the request, and the 3 best candidates are selected. 
All selected candidates are then concatenated and appended to the prompt which is then passed to the \textit{Composer LLM}. 
The \textit{Retrieval LLM}'s own system prompt instructs it on what tags are present in the database and how it should format the tags. 
To keep costs low\footnote{The bill for the OpenAI API use for this project amounted to $\$7.17$}, we use \texttt{ChatGPT-3}\cite{Introducing} as our \textit{Retrieval LLM}.

\noindent \textbf{Composer LLM}
We use \texttt{ChatGPT-4}\cite{Introducing,openai2023gpt4} as our \textit{Composer LLM}.
The system prompt for this LLM contains detailed instructions on the system's role and how to format its output. 
We also ask the LLM to write a short commentary of the piece it will generate prior to generation. We do this to get the LLM to develop a chain-of-thought which has shown to improve performance in other tasks\cite{wei2022chain}.

\section{Examples}\label{sec:examples}
We now present three examples of interactions with the system, including commentary of each output. All outputs have been cross-checked against the session database for exact copies.
\subsection{Conditional Generation}
\begin{figure}[h]
\begin{mdframed}[innerleftmargin=5pt, innerrightmargin=5pt]
\footnotesize
\texttt{%
\textbf{Prompt:} Generate a piece of irish folk music.\\
\textbf{Composer comment:} This will be a lively Irish reel in 4/4 time with a major key of D. The A-part of the tune will repeat twice, as will the B-part.\\
\textbf{Generated tune:\\}
}

\includegraphics[width=\linewidth]{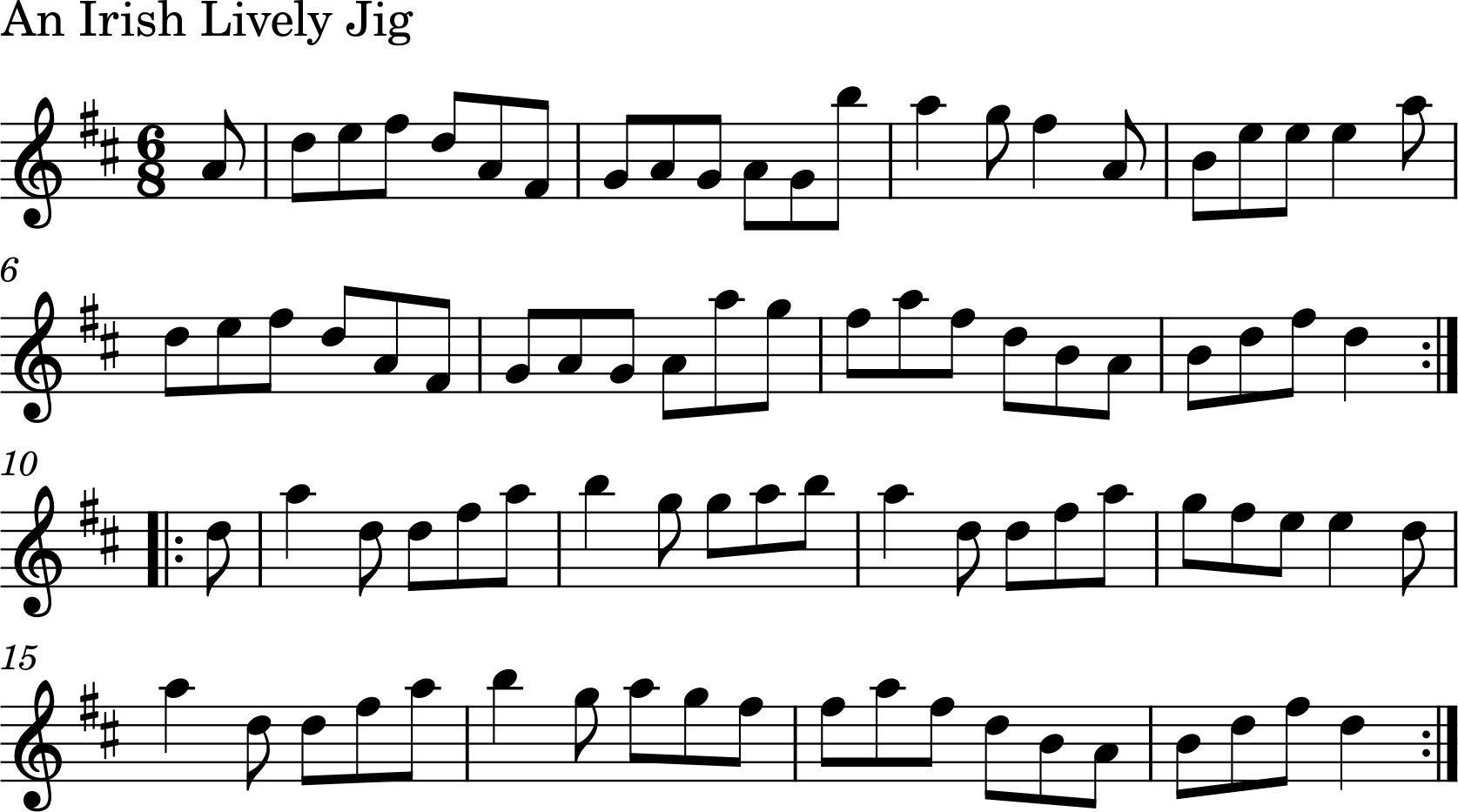}
\end{mdframed}
\label{fig:jig}
\caption{Prompting for Irish folk music resulted in a jig with recognizable style.}
\end{figure}

The jig in Figure \ref{fig:jig} follows the traditional 6/8 rhythm.
Both parts feature 2- or 4-bar long melodic units where material is reused, and they share the last two bars tying the tune together.
The jumps in the A-section are the only strange sounding aspect.

\subsection{Style Translation}
\begin{figure}[h]
\begin{mdframed}[innerleftmargin=5pt, innerrightmargin=5pt]
\footnotesize
\texttt{%
\textbf{Prompt:} Turn this tune into a polka T:An Irish Lively Jig M:6/8 K:Dmajor A|def $...$ d2:|\\
\textbf{Composer comment:} Converting a jig into a polka mainly involves changing from 6/8 to 2/4 time and shortening the notes accordingly.\\
\textbf{Generated tune:}\\
}

\includegraphics[width=\linewidth]{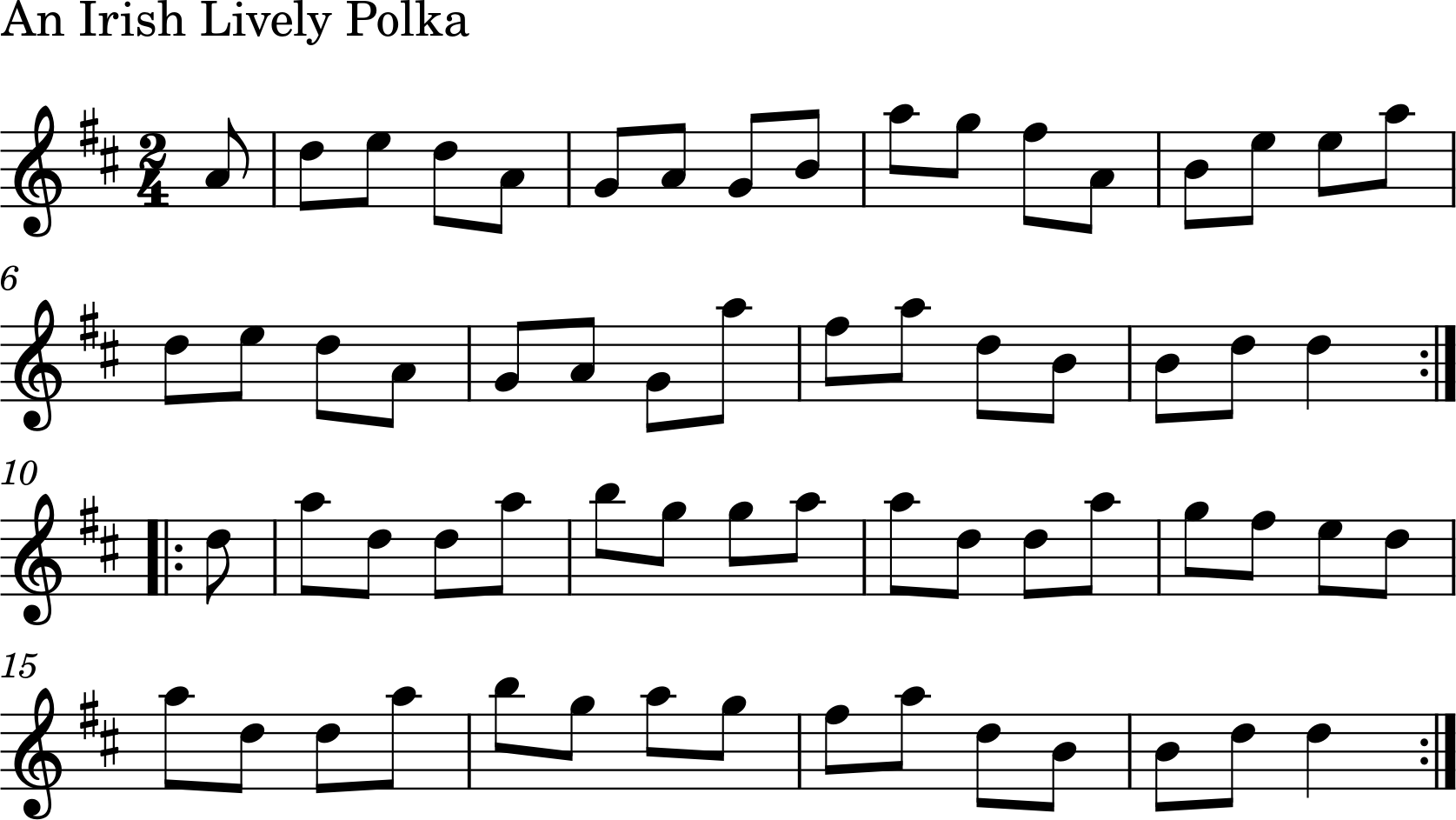}
\end{mdframed}
\caption{Converting the previous jig into a polka.}
\label{fig:polka}
\end{figure}

In Figure \ref{fig:polka} we have the model turn the previous tune into a Polka. This entails changing a ternary rhythm into a binary one.
In the output, we still see a recognizable melodic contour and voice leading, although the pickup measures are not correct anymore.

\subsection{Tune Completion}
\begin{figure}[h]
\begin{mdframed}[innerleftmargin=5pt, innerrightmargin=5pt]
\footnotesize
\texttt{%
\textbf{Prompt:} Write a beginning for this ending: |:A|dAF DAF|GAG Aag|dAF DAF|Ffe e2a|dAF DAF|GAG Aag|faf dBA|Bdf d2:|\\
\textbf{Composer comment:} This tune begins in D major, with a lively rhythm and melodies that mirror the ending's pattern for consistency\\
\textbf{Generated tune:}\\
}

\includegraphics[width=\linewidth]{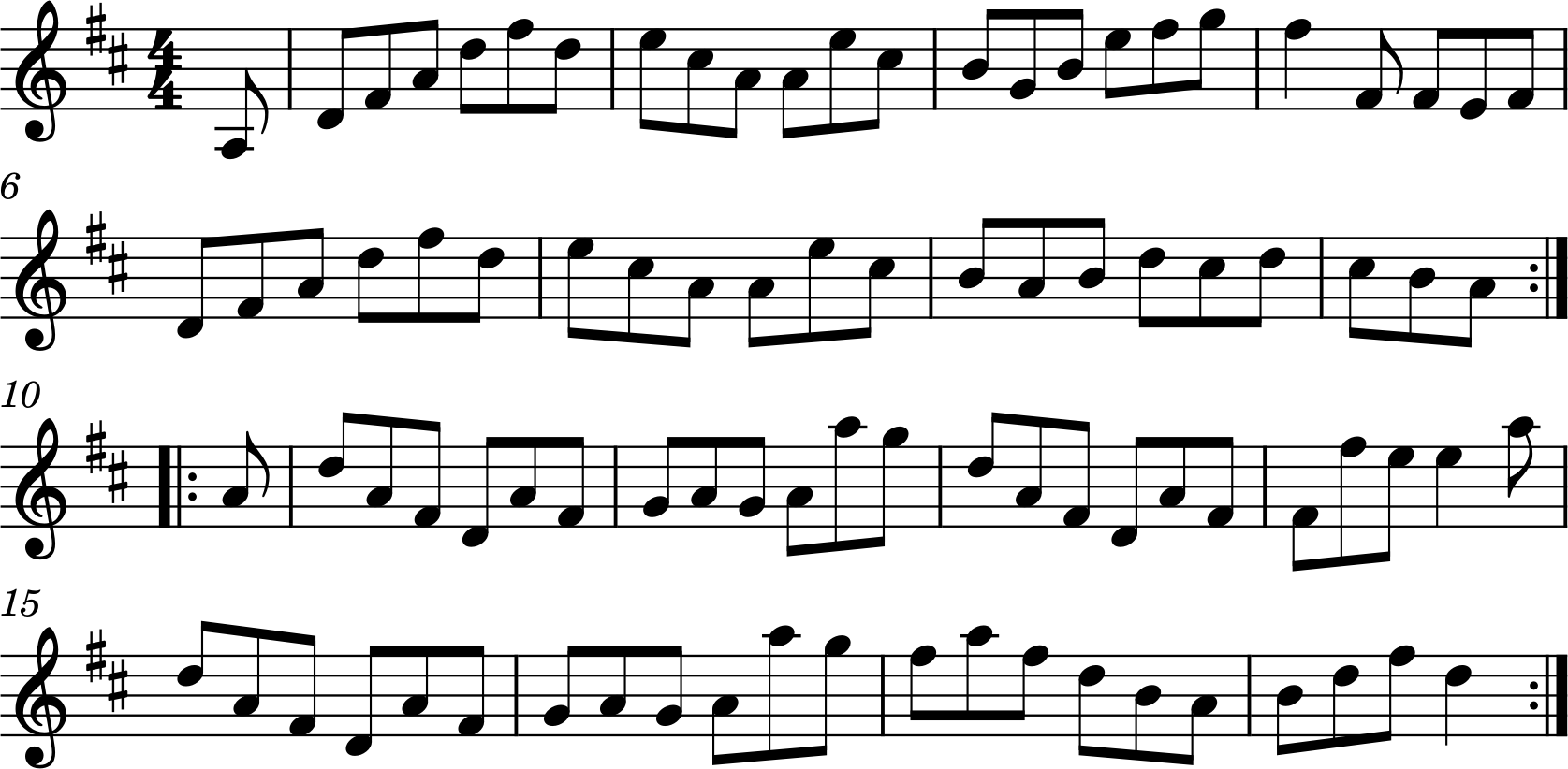}
\end{mdframed}
\caption{Adding an A-section to the input.}
\label{fig:beginning}
\end{figure}

In Figure \ref{fig:beginning} a new A-section is created given the prompt.
The output seems to mirror the melodic motion in the input, creating a nice contrast. 
The final bar lacks 2/8, but this is a tolerable mistake we also see in other folk models \cite{ijcai2022p681,Sturm2016a}.

\section{Future work}

While our preliminary experiments show that our system can address many musical tasks, we plan to further develop the application of LLMs to the analysis and generation of symbolic music.  We also want to evaluate the generation quality of the proposed system and compare it to previous abc music generation approaches.\cite{Sturm2016a,ijcai2022p681} 

\newpage

\section{Acknowledgments}

This paper is an outcome of MUSAiC, a project that has
received funding from the European Research Council
under the European Union’s Horizon 2020 research and
innovation program (Grant agreement No. 864189).

\bibliography{ISMIRtemplate}

\end{document}